\documentstyle[11pt,epsfig]{article}

\def\laq{~\raise 0.4ex\hbox{$<$}\kern -0.8em\lower 0.62
ex\hbox{$\sim$}~}
\def\gaq{~\raise 0.4ex\hbox{$>$}\kern -0.7em\lower 0.62
ex\hbox{$\sim$}~}

\def\beq{\begin{equation}}
\def\eeq{\end{equation}}
\def\bea{\begin{eqnarray}}
\def\eea{\end{eqnarray}}
\def\bean{\begin{eqnarray*}}
\def\eean{\end{eqnarray*}}

\def \tb {{\overline {t}}}

\def\l {\langle}
\def\re {\rangle}

\def \pa {\partial}
\def \ra {\rightarrow}
\def \ti {\widetilde}
\def \ha {\widehat}

\def \Da {\Delta}
\def \da {\delta}
\def \b {\beta}
\def \a {\alpha}

\def \ga {\gamma}

\def \Sg {\Sigma}
\def \da {\delta}
\def \ep {\epsilon}

\def \Om {\Omega}
\def \noi {\noindent}

  \def\be{\begin{equation}}
    \def\ee{\end{equation}}
    \def\ba{\begin{eqnarray}}
    \def\ea{\end{eqnarray}}

\begin{document}

\begin{titlepage}

\begin{flushright}
BA-TH/602-09\\
CERN-PH-TH/2009-010\\
arXiv:0901.1303
\end{flushright}

\vspace{0.5cm}

\begin{center}

\huge
{Gauge invariant averages \\ for   the cosmological backreaction}

\vspace{0.8cm}

\large{M. Gasperini$^{1,2}$, G. Marozzi$^{3,4}$ and G. Veneziano$^{5,6,7}$}

\normalsize

\vspace{0.5cm}

{\sl $^1$Dipartimento di Fisica, Universit\`a di Bari, \\
Via G. Amendola 173, 70126 Bari, Italy}

\vspace{.1in}

{\sl $^2$Istituto Nazionale di Fisica Nucleare, Sezione di Bari, Bari, Italy}

\vspace{.1in}{\sl $^3$Dipartimento di Fisica, Universit\`a di Bologna, \\
Via Irnerio 46, 40126 Bologna, Italy}

\vspace{.1in}

{\sl $^4$Istituto Nazionale di Fisica Nucleare, Sezione di Bologna, Bologna,
Italy}

\vspace{.1in}

{\sl $^5$ Coll\`ege de France, 3, rue d'Ulm, 75005 Paris, France }

\vspace{.1in}

{\sl $^6$CERN, Theory Unit, Physics Department, \\ CH-1211 Geneva 23,
Switzerland}

\vspace{.1in}

{\sl $^7$Galileo Galilei Institute for Theoretical Physics, Arcetri, Italy, \\ and Istituto Nazionale di Fisica Nucleare, Sezione di Firenze, Italy}

\vspace*{1cm}

\begin{abstract}
\noi
We show how  to provide suitable gauge invariant prescriptions for the classical spatial averages (resp. quantum expectation values)  that are needed in the evaluation of  classical (resp. quantum) backreaction effects.  We also present examples illustrating how the use of gauge invariant prescriptions can avoid interpretation problems and prevent  misleading conclusions. 
\end{abstract}

\end{center}

\end{titlepage}

\newpage

\parskip 0.2cm

\section{Introduction}
\label{sec1}
\setcounter{equation}{0}

The study of the possible dynamical influence of (small) inhomogeneities on the large-scale evolution of a cosmological background has recently attracted considerable interest, from both a  theoretical and  a phenomenological point of view. Such a study is of particular relevance for the case of inflationary backgrounds, where macroscopic inhomogeneities necessarily arise as a consequence of  the inflationary amplification of the quantum fluctuations of the metric and of the matter sources. In that case, the amplified perturbations may become large enough to eventually modify the initial cosmological evolution, triggering in this way a true backreaction mechanism (as first pointed out and discussed in \cite{1}). But inhomogeneities could also affect in a non-trivial way  the present cosmological evolution, as suggested by recent interpretations of the dark-energy source as the ``backreaction effect'' of appropriately smoothed-out inhomogeneities \cite{2} (see however \cite{2a}). 

In order to take into account the influence of such a backreaction on the large-scale geometry what is needed, in particular, is the application of a well defined averaging procedure for smoothing-out  the perturbed (non-homogeneous) geometric parameters. The averaging may be referred to a suitably chosen hypersurface, such as the one comoving with the gravitational sources (as in \cite{3}), or one of constant energy density, or one of constant curvature, etc: the choice will depend, in general, on the specific problem under study. What is also needed is a computation of the non-homogeneous parameters up to the second perturbative order, at least when the average of the first-order perturbations is vanishing: this is indeed what happens in the case of the quantum fluctuations, whose statistical distribution is expected to keep its original stochastic properties even after the inflationary amplification. 

The computations of the non-homogeneous corrections to a perturbed cosmological background, on the other hand, are affected by a 
well-known ambiguity due to the possible choice of different ``gauges'', i.e. of different possible parametrizations of the non-homogeneous degrees of freedom, connected among them by coordinate transformations (see e.g. \cite{4}). The problem thus arises
of using averaging prescriptions which -- quite independently of the hypersurface (or otherwise-defined portion of space-time) chosen for the smearing of perturbations -- turn out to be independent of the particular parametrization, thus providing a result which is gauge invariant, at least to second order (see \cite{5} for a recent attempt in this direction, focused on the particular case of the de Sitter geometry). The aim of this paper is to propose a covariant average prescription satisfying the above property and briefly mention some related applications to primordial cosmological backgrounds. 
Even though our approach is in principle non-perturbative, in this paper, 
for technical reasons,  we do not develop it explicitly beyond second-order.
A detailed study of the implications of our proposal for estimating backreaction effects in a variety of cosmological setups will be presented elsewhere.

The rest of the paper is organized as follows. In Sect. \ref{sec2}, after stressing the difference between coordinate reparametrizations and gauge transformations, we discuss on general grounds the gauge transformation properties of four-dimensional volume integrals. In Sect. \ref{sec3} we concentrate on the quasi-homogeneous cosmological case, introduce a generally covariant and 
gauge-invariant averaging prescription for non-homogeneous scalar variables, and compute its explicit expression -- up to second order in the perturbative expansion of all background inhomogeneities -- for both spatial averages of classical perturbations and expectation values of quantized perturbations. Possible examples and applications of these results to the computation of backreaction effects are briefly mentioned in Sect. \ref{sec4}. Our conclusions are finally summarized in Sect. \ref{sec5}.

\section {Gauge (non)-invariance of space-time integrals}
\label{sec2}
\setcounter{equation}{0}

In the context of a smoothed space-time geometry, the computation of the ``effective'' averaged parameters relies on the integration of their inhomogeneities over an appropriate space-time domain (see e.g. \cite{6}). The behaviour of such averaged parameters under gauge-transformations is thus determined by the transformation properties of four-dimensional integrals defined over the space-time manifold. In order to discuss this point, let us first recall the (slightly subtle) difference between general coordinate transformations (GCT) and gauge transformations (GT). 

Let us consider a (typically non-homogeneous) scalar field $S(x)$  defined on a generic four-dimensional space-time manifold ${\cal M}_4$ equipped with a Riemannian metric $g_{\mu \nu}$ and parameterized by a given set of coordinates $x^\mu$. Under a GCT:
\beq
x \rightarrow \ti{x} = f (x), ~~~ ~~~~~~~~ x = f^{-1}(\ti{x}),
\label{21}
\eeq
$S$ transforms, by definition, as $S(x) \rightarrow \ti{S}(\ti x)$, where:
\beq
\label{22}
\ti{S}(\ti x)  = S(x).
\eeq
Under the associated GT (or local field reparametrization) -- where, by definition, old and new fields are evaluated at the same space-time point $x$ -- $S$ transforms as $S(x) \rightarrow \ti{S}(x)$, where:
\beq
\label{23}
\ti{S}(x)  = S(f^{-1}(x)).
\eeq
Note that this definition applies not only to infinitesimal, but also to {\em finite} coordinate transformations, and that the standard GCT relation (\ref{22}) is immediately recovered -- using Eq. (\ref{21}) -- when the above equation is evaluated at the point $\ti x$. In general, $\ti S(x) \ne S(x)$, namely a scalar field {\em is not}  invariant under a general GT.

For  the metric tensor the GT associated with Eq. (\ref{21}) is similarly defined as $g_{\mu\nu}(x) \ra \ti g_{\mu\nu}(x)$, where:
\beq
\ti g_{\mu\nu}(x) = \left[ {\pa x^\a\over \pa f^\mu}
 {\pa x^\b\over \pa f^\nu}\right]_{f^{-1}(x)} g_{\a\b}(f^{-1}(x)).
 \label{24}
\eeq
The subscript appended to the square brackets denotes that, after computing the derivatives of the inverse Jacobian matrix $(\pa f /\pa x)^{-1}$, the variable $x$ has to be replaced by $f^{-1}(x)$. 
Consequently, for $g\equiv \det g_{\mu\nu}$ the GT transformation reads $\sqrt{-g(x)} \rightarrow
\sqrt{- \ti{g}(x)}$, where: 
\beq
\sqrt{- \ti{g}(x)}  = \left|\pa x \over \pa f\right|_{f^{-1}(x)}
\sqrt{-g(f^{-1}(x) )} \, ,
\label{25}
\eeq
with $\left|\pa x / \pa f\right|$ the inverse Jacobian determinant associated with the GCT (\ref{21}). 

Consider now the space-time integral of the scalar $S$ over a four-dimensional region $\Om$ of ${\cal M}_4$, spanned by the coordinates $x^\mu$, and define:
\beq
F(S,\Om) = \int_{\Om(x)} d^{4} x  \sqrt{-g(x)} \,S(x). 
\label{26}
\eeq
Let us now compute its transformation properties under the associated GT that keeps $x$ unchanged. Assuming that the region $\Om$ also stays the same:
\beq
\label{27}
F(S,\Om)  \rightarrow \ti{F}(\ti{S}, \Om) = \int_{\Om(x)} d^{4 }x \sqrt{-\ti{g}(x)} \,\ti{S}(x), 
\eeq
where $\ti S$ and $\ti g$ are given by Eqs. (\ref{23}), (\ref{25}), respectively. We now perform a change of integration variable from $x$ to $\ha x = f^{-1}(x)$ (note that $\ha x \ne \ti x$\,!)
  to obtain: 
 \beq
 \ti{F}(\ti{S}, \Om) = \int_{\Om(f(\ha x))} d^{4 }\ha x 
 \left|\pa f\over \pa \ha x \right|_{\ha x}
 \sqrt{-\ti{g}(f(\ha x))} \,\ti{S}(f(\ha x)).
 \label{29}
 \eeq 
Using Eqs. (\ref{23}), (\ref{25}), we are finally led to 
 \beq
 \ti{F}(\ti{S}, \Om) = \int_{\ha\Om(\ha x)} d^{4 }\ha x 
 \sqrt{-{g}(\ha x)} \,{S}(\ha x) = F(S, \ha\Om) \, ,
 \label{210}
 \eeq 
 where we have introduced the transformed region $\ha \Om$, defined by $\ha \Om (\ha x) \equiv \Om(f(\ha x))$. 

Clearly the gauge-transformed integral (\ref{210}) and the original one (\ref{26}) differ since $\ha \Om \ne \Om$. Hence, the expression (\ref{26}) is not gauge invariant if we assume -- as we did -- that the integration  region is left unaffected by the GT. 

This general result suggests a direct way to obtain gauge-invariant integrals, simply by anchoring the definition of the integration region $\Om$ to the gauge choice so  that, under a GT, the region $\Om$ automatically transforms in such a way as to exactly compensate the final variation $\ha \Om -\Om$. 
This can be done as follows: define the integration region $\Om$ of ${\cal M}_4$ in terms of a  window function $W_\Om$,  so that Eq. (\ref{26}) can be rewritten as
\beq
F(S,\Om) = \int_{\Om(x)} d^{4} x  \sqrt{-g(x)} \,S(x) \equiv
 \int_{{\cal M}_4} d^{4} x  \sqrt{-g(x)} \,S(x)W_\Om(x),
\label{211}
\eeq
and assume that $W_\Om(x)$, rather than being a function given from the outside, is a scalar function built out of the dynamical fields at our disposal
(explicit examples of step-like and delta-like window functions will be discussed below and in the next section). 
In this case, under a GT:
\beq
W_\Om(x) \ra \ti W_\Om(x) = W_\Om(f^{-1}(x)),
\label{212}
\eeq
and the integration region is consequently transformed from $\Om(x)$ to $\Om(f^{-1}(x))$. If we now repeat the above procedure  (performing a GT and a change of integration variable), starting however from the new definition (\ref{211}), and using (\ref{212}), we arrive at the result
 \beq
 \ti{F}(\ti{S}, \Om) =  \int_{{\cal M}_4} d^{4 }\ha x 
 \sqrt{-{g}(\ha x)} \,{S}(\ha x) W_\Om(\ha x)
 = \int_{\Om(\ha x)} d^{4 }\ha x 
 \sqrt{-{g}(\ha x)} \,{S}(\ha x) \equiv  {F}({S}, \Om). 
 \label{213}
 \eeq 
The integral (\ref{211}) is thus automatically gauge invariant as a
consequence of its invariance under GCTs, with the appropriate
``gauge-shifting'' of the integration region under the GT. 

Let us conclude this section with an explicit example illustrating how the breaking of gauge invariance (and general covariance) is related to the transformations properties of the boundary of the integration region. We shall consider a step-like window function, selecting a 
cylinder-like  region $\Om$  of ${\cal M}_4$, with temporal boundaries determined by the two space-like hypersurfaces on which a suitable scalar field $A(x)$ (with time-like gradient $\pa_\mu A$) assumes the constant values $A_1$ and $A_2$; the region is bounded in space by the coordinate condition $B(x) < r_0$, where $B(x)$ is a suitable (positive) function of the coordinates with space-like gradient $\pa_\mu B$, and $r_0$ is a positive constant. More explicitly, we define
\beq
 W_\Om(x) = \theta(A(x)- A_{1}) \theta(A_{2}- A(x))\theta(r_0-B(x))\, .
\label{theta1}
\eeq
where $\theta$ is the Heaviside step function. 

Using this window in the definition (\ref{211}), and applying  a
GT, we find that the temporal boundaries of $\Om$ are changed because of the gauge shifting $A(x) \ra A(f^{-1}(x))$). The transformation of the spatial boundary is similarly controlled by the transformation properties of $B(x)$, If $B$ is a scalar we have  $B(x) \ra B(f^{-1}(x))$), the property (\ref{212}) is then satisfied, and the integral is gauge invariant. However, for the cosmological backgrounds which are of central interest to this paper, all fields are naturally of quasi-homogeneous type, and their gradients are typically time-like. In such a context we cannot covariantly define the spatial boundaries as done for the temporal ones, for lack of appropriate fields at our disposal. Using for $B(x)$ a function of the coordinates, which does not change under a GT, the integral (\ref{211}) fails to be gauge invariant: repeating the same procedure as before we find that the 
final result (\ref{213}),  is replaced in this case by: 
 \beq
 \ti{F}(\ti{S}, \Om) - {F}({S}, \Om) =
  \int_{{\cal M}_4} d^{4 } x 
 \sqrt{-{g}( x)} \,{S}( x) \Da W_\Om( x), 
 \label{215}
 \eeq 
where
\be
\Da W_\Om(x)= \theta(A(x)-A_1)\theta(A_2-A(x))
\left[\theta (r_0-B(f(x)))-
\theta (r_0-B(x)) \right].
\label{216b}
\ee
We see that the breaking of gauge invariance is given by an integral over a ring-shaped  region  lying near the space  boundary, $B= r_0$, and having a ``thickness'' $\Delta r$ controlled by the magnitude of the GT. 

We shall now argue that this breaking term tends to be subdominant for large enough 
spatial volumes. The gauge transformations we are interested in are related to the inhomogeneities themselves, which are assumed to be a small perturbation of the cosmological background. We may then expand the coordinate transformation with respect to the first-order parameter $\ep^\mu$ as
\be
x^\mu \rightarrow \tilde{x}^\mu= x^\mu + \epsilon^\mu(x) + \dots
\label{4}
\ee
(our considerations can be extended to second order in the perturbations without altering the general conclusions). 
Inserting  the Taylor expansion of Eq. (\ref{216b})  in Eq. (\ref{215}) 
we obtain:
\bea
 & & ~~~~~~~~\ti{F}(\ti{S}, \Om) - {F}({S}, \Om)=\nonumber \\
& & = - \int d^{4}x \sqrt{-g} \, S(x) \theta(A(x)- A_{1}) 
\theta(A_{2}- A(x)) \delta\left(r_0-B(x)\right)
\frac{\partial B(x)}{\partial
x^\mu}\epsilon^\mu, 
\label{219b}
\eea
and we note that the relevant quantity to be considered is 
the relative magnitude of the breaking term, namely:
\be
\frac {\Delta F}{F} \equiv \frac{ \ti{F}(\ti{S}, \Om) - {F}({S}, \Om)}{{F}({S}, \Om)}.
\ee
In order to estimate the above expression we can use coordinates in which $A(x)$ and $B(x)$ are identified with the time and radial coordinates $t$ and $r$. The remaining  coordinates will be like the two angular coordinates ($\theta$ and $\phi$) on the 2-sphere. We then get:
\be
\frac {\Delta F}{F}  = \left( \int_{A_1}^{A_2}dt \int d\theta d\phi 
 \sqrt{-g(x_0)} \, S(x_0) \epsilon^r(x_0) \right)  \left(\int_{A_1}^{A_2}dt \int d\theta d\phi  \int_0^{r_0} dr 
\sqrt{-g(x)} \, S(x)\right)^{-1}
\ee
where $(x_0)$ stands for $(t, r_0,\theta, \phi)$, and $\ep^r$ is the radial gauge parameter of the transformation (\ref{4}). 

 Barring cancellations, we see that this quantity is of order $\epsilon^r(x_0)/d(r_0)$, where $d(r_0)$ is the proper size of the spatial region of integration,  and $\epsilon^r$ is determined by some generic perturbation $\delta X={\cal L}_{\epsilon} X$, where ${\cal L}_{\epsilon}$ denotes the Lie derivative, through:
\be
  \epsilon^r \;\frac{\partial X}{X} \sim \frac{\delta X}{X} \,.
\ee
This means that $\epsilon^r$ is related to the perturbation ${\delta X}/{X}$ through a dimensionful factor $\lambda$ (possibly controlled   either by the characteristic time scale of the background or by the characteristic wavelength of the inhomogeneity). In any case, we expect ${\Delta F}/{F}$ to go to zero like the ratio of such $\lambda$ to the proper size of the region considered, and thus to vanish in the limit of an infinite spatial volume.

\section {A gauge-invariant prescription for spatial averages \\ and quantum expectation values}
\label{sec3}
\setcounter{equation}{0}

We have shown, in the previous section, that four-dimensional integrals can be made exactly gauge invariant under general (and finite) coordinate transformations by an appropriate definition of the integration domain. We have also briefly discussed how the breaking of gauge invariance may be related to the choice of the spatial boundaries of integration. Our main task, in the present section, will be the application of the above results to the integrals which are of central interest to this paper, namely to the average integrals arising in the computation of the cosmological backreaction. 

Depending on the context in which the backreaction is approached, there are two types of averaging procedure: spatial (or {\em ensemble}) average of classical variables, and (vacuum) expectation values of quantized fields. In both cases, the associated integrals (and the results of the averaging procedure) are not gauge invariant, in general (see e.g. \cite{7,8,9}, as well as the examples of Sect. \ref{sec4}), and one has to face the problem of transforming and physically re-interpreting the results from one gauge to another. In order to solve this problem we are led to the question: is it possible to define gauge-invariant (spatial averaging) integrals?

To answer this question we first observe that expectation values of quantum 
operators can be extensively interpreted (and re-written) as spatial integrals 
weighted by the integration volume $V$, according to the general prescription
\beq
\l \dots \re  ~~~\ra ~~~ V^{-1} \int_V d^3 x \left( \dots\right)\,,
\label{31}
\eeq
where, at least for the cosmological backgrounds we are interested in, the integration volume extends to all three-dimensional space.

This clearly associates expectation values with classical volume averages. Spatial volume averages, on the other hand, can be covariantly obtained from the four-dimensional (and possibly gauge-invariant) integrals discussed in the previous section simply by using a delta-like window function, so as to confine the integration on the spatial hypersurface chosen to perform the averaging procedure. 

Such a hypersurface can be selected, as before, as the one on which a suitable scalar field $A(x)$ (with time-like  gradient $\pa_\mu A$) takes the constant value $A_0$. We can also specify the range of integration across this hypersurface by inserting, into the window function, a step-like definition of the spatial boundary: for instance, we can impose on a suitable (positive) function of the coordinates $B(x)$
 (with space-like  gradient $\pa_\mu B$) to be bounded by the constant positive value $r_0>0$. This leads to 
 \beq
 W_\Om(x)= \da (A(x)-A_0) \theta (r_0-B(x)),
 \label{32}
 \eeq
 where the delta function localization may also be seen as the limit $A_2 \ra A_1=A_0$ in which the ``thickness'' of the step-like time slicing of Eq. (\ref{theta1}) is sent to zero. 
 
 Using the above window, we are eventually led to define the averaging prescription for a scalar object $S$ as 
 \beq
 \l S\re_{\{A_0,r_0\}} = {F(S, \Om)\over F(1, \Om)}= 
{\int_{{\cal M}_4} d^4x \sqrt{-g} \, S \,\delta(A-A_0)\, \theta(r_0-B) 
\over \int_{{\cal M}_4}  d^4x \sqrt{-g} \,\delta(A-A_0) \,\theta(r_0-B)}. 
\label{33}
\eeq
We recall (and stress) that the two integrals appearing in this definition are both either gauge invariant or not, depending on the transformation properties of the spatial boundary determined by $B(x)$. In particular, they are covariant and gauge invariant if $B$ transforms under GT as a scalar, while they are not if the definition of the boundary is independent of the gauge, as discussed in the previous section. 

In order to explicitly rewrite the above definition in terms of spatial (three-dimensional) integrals we can now exploit the properties of the delta function, and perform the time integration. Consider, to this purpose, the change of integration variable from $t$ to $\tb$, defined by $t= h(\tb,x)$, where the function $h$ is chosen so as to make the scalar field $A$ homogeneous, i.e.
\beq
A(h(\tb,x),x)= \overline A (\tb, x) \equiv A^{(0)}(\tb).
\label{34}
\eeq
After this change of variable the average (\ref{33}) becomes 
\beq
\langle S \rangle_{\{A_0,r_0\}}= 
{\int_{{\cal M}_4} d \tb d^3x  \sqrt{-\overline{g}(\tb, {x})} \,
  \overline{S}(\tb, {x}) \, \delta(A^{(0)}(\tb)-A_0)\,\theta(r_0-B(h(\tb, {x})
, {x})) 
\over \int_{{\cal M}_4} d \tb d^3x  \sqrt{-\overline{g}(\tb, {x})} \,  
\delta(A^{(0)}(\tb)-A_0) \,
\theta(r_0-B(h(\tb, {x}), {x}))},
\label{35}
\eeq
where we have used the scalar transformation property $S(h(\tb,x),x)= \overline S (\tb,x)$, and we have absorbed the Jacobian factor $|\pa t/\pa \tb|$ into the definition of $\overline g(\tb, x)$, according to Eq. (\ref{25}). The delta-function integration can now be easily performed, and gives
\beq
\langle S \rangle_{\{A_0,r_0\}}= 
{\int_{{\Sg}_{A_0}} d^3x  \sqrt{-\overline{g}(t_0, {x})} 
\,~ \overline{S}(t_0, {x}) \,\theta(r_0-B(h(t_0, x), {x})) 
\over \int_{{\Sg}_{A_0}}  d^3x  \sqrt{-\overline{g}(t_0, {x})} \,  
\theta(r_0-B(h(t_0, x), {x}))},
\label{36}
\eeq
where we have called $t_0$ the time $\tb$ when $A^{(0)}(\tb)$ takes the constant values $A_0$. The suffix $\Sg_{A_0}$ on the integral is  there to recall that we are averaging a scalar object $S$ on a section of the three-dimensional 
hypersurface $\Sg_{A_0}$,  where the given scalar field 
$A(x)$  takes the constant values $A_0$. 

The gauge invariance of this expression is determined, as in the four-dimensional case, by the transformation properties of the spatial boundary, controlled by $B(x)$. We have already stressed that in the case of interest to this paper -- i.e. for quasi-homogeneous cosmological backgrounds -- a natural candidate for the scalar field $B(x)$ with space-like gradient is missing. Using for $B(x)$ a function of the coordinates which does not change under a GT, and following the discussion of Sect. \ref{sec2}, it is clear that the above integrals will be gauge invariant only in the limit of an infinite spatial volume.
So, to obtain a gauge-invariant average prescription, we must consider the limit in which $r_0 \ra \infty$ 
(or $B \ra 0$), and the integration extends over all three-dimensional space. 
In this limit the step-like boundary disappears, and we obtain 
\beq
\langle S \rangle_{A_0}= 
{\int_{{\Sg}_{A_0}} d^3x  \sqrt{-\overline{g}(t_0, {x})} \,~ \overline{S}(t_0, {x})
\over \int_{{\Sg}_{A_0}}  d^3x  \sqrt{-\overline{g}(t_0, {x})}}.
\label{37}
\eeq

This represents our gauge invariant prescription for the average of a scalar object $S$ on the hypersurface $\Sg_{A_0}$, for any
given scalar field $A(x)$ determining a foliation of space-time into spatial hypersurfaces $\Sg_A$ with $A=$ const. 
We should note, in this generalized gauge-invariant definition of the average of $S$, the presence under 
the integral not of $S$ but of $\overline S$, i.e. of the variable $S$ transformed to the 
coordinate frame in which $A(x)$ is homogeneous.

It should be noted, also, that Eq. (\ref{37}) directly generalizes both classical volume averages and quantum expectation values. In the quantum case, in particular, Eq. (\ref{37}) can be rewritten as
\beq
\langle S \rangle_{A_0}={\l   \sqrt{-\overline{g}(t_0, {x})} 
 \,~ \overline{S}(t_0, {x}) \re \over  \l   \sqrt{-\overline{g}(t_0, {x})} \re},
 \label{38}
 \eeq
where the brackets without suffix, on the right-hand side of the equality, refer to the standard operatorial prescription  (\ref{31}) with integration over all three-dimensional space. We stress that the two entries of this ratio are not separately gauge invariant, but the ratio itself -- equivalent to Eq. (\ref{37}) -- is indeed invariant, as can be directly verified by applying the above definitions (see also the explicit examples of Sect. \ref{sec4}). 
 
 For a direct check of gauge invariance, as well as for later applications to quasi homogeneous cosmological backgrounds, we will present here an explicit expansion (up to second order) of the generalized average $\langle S \rangle_{A_0}$ in terms of  conventional averages. The subsequent discussion can be applied to both the quantum expectation value (\ref{38}) and its classical analogue (\ref{37}), for the average of a quasi-homogeneous scalar variable $S(x)$ whose inhomogeneities can be perturbatively expanded as
 \beq
 S= S^{(0)}+ S^{(1)}+ S^{(2)} + \dots
 \label{39}
 \eeq
Here $S^{(0)}$ is the dominant homogeneous part associated to the unperturbed background, $S^{(1)}$ contains the first-order non-homogeneous corrections, and so on. A similar expansion is assumed to apply also to to other relevant variables $g_{\mu\nu}$ and $A$. We will also assume that the conventional average of all first-order terms is vanishing ($\l S^{(1)}\re=0$, $\l g^{(1)}\re=0$, etc \dots), as is indeed the case of inhomogeneities of quantum origin (like those arising from the amplification of the vacuum fluctuations in the context of inflationary backgrounds). 

By expanding Eq. (\ref{38}) up to second order, under the above assumptions, we first obtain the result 
\beq
\langle S \rangle_{A_0} = S^{(0)} +\langle \overline{S}^{(2)} \rangle +
\frac{1}{(\sqrt{-g})^{(0)}}\langle \overline{S}^{(1)}(\sqrt{-\overline{g}})^{(1)}\rangle.
\label{310}
\eeq
We have then to express the transformed (barred) fields, relative to the frame in which $A$ is homogeneous, in terms of the original (unbarred) fields, in a general gauge. To this purpose let us first consider the ``infinitesimal'' coordinate transformation parametrized by the  first-order, $\ep_{(1)}^\mu$, 
and second-order, $\ep_{(2)}^\mu$,  generators as \cite{9bis}
\beq
x^\mu \rightarrow \tilde{x}^\mu= x^\mu + \epsilon^\mu_{(1)} +\frac{1}{2}
\left(\epsilon^{\nu}_{(1)}\pa_\nu \epsilon^{\mu}_{(1)} + \epsilon^{\mu}_{(2)}\right) + \dots
\label{311}
\eeq
where
\beq
\ep_{(1)}^\mu= \left( \ep_{(1)}^0, \pa^i \ep_{(1)}+ \ep_{(1)}^i 
\right), ~~~~~~~~~
\ep_{(2)}^\mu= \left( \ep_{(2)}^0, \pa^i \ep_{(2)}+ \ep_{(2)}^i \right) 
\label{312}
\eeq
(we have explicitly separated the scalar part from the pure transverse vector part $\ep_{(1)}^i$, $\ep_{(2)}^i$, see e.g. \cite{4}). Let us recall, also, that the associated gauge transformation of the scalar field $S$
is, to first order, 
\be
S^{(1)}~~ \rightarrow ~~\ti{S}^{(1)}=S^{(1)}-\ep_{(1)}^0 \dot{S}^{(0)},
\label{315}
\ee
and, to second order, 
\bea
S^{(2)} ~~\rightarrow &&~~\ti{S}^{(2)}=  S^{(2)}-\ep_{(1)}^0 \dot{S}^{(1)}
-\left(\ep_{(1)}^i+\partial^i \ep_{(1)}\right) \partial_i S^{(1)}
\nonumber \\ 
&&+\frac{1}{2}\left[\ep_{(1)}^0 
\partial_t (\ep_{(1)}^0 \dot{S}^{(0)})+ \left(\ep_{(1)}^i+\partial^i \ep_{(1)}\right)\partial_i
\ep_{(1)}^0 \dot{S}^{(0)}-\ep_{(2)}^0 \dot{S}^{(0)}\right],
\label{316}
\eea 
where dots denote time derivatives, i.e. $\dot S= \pa_tS$. 
Similarly, for the first-order transformations of $\sqrt{-g}$, using the tensor transformation of the metric (or expanding Eq. (\ref{25}) up to first order) we have
\beq
(\sqrt{-g})^{(1)} ~\rightarrow ~(\sqrt{-\ti{g}})^{(1)}
=(\sqrt{-g})^{(1)}+(\sqrt{-g})^{(0)}\left[-
\dot{\epsilon}_{(1)}^0-\nabla^2 \epsilon_{(1)}-\partial_t\left(\ln{(\sqrt{-g})^{(0)}}\right)\epsilon_{(1)}^0 \right]\,.
\label{317}
\eeq

We can now apply the above general transformations of $S$ and $\sqrt{-g}$ to the particular case of the transformation $t \ra \tb$, used to make $A$ homogeneous (see Eq. (\ref{34})). To this purpose we first expand to second order the transformation $t= h(\tb,x)$, and invert it, so as to recast it in the general form (\ref{311}), namely: 
\beq
\tb= t+ \epsilon^0_{(1)} +\frac{1}{2}
\left(\epsilon^{\nu}_{(1)}\pa_\nu \epsilon^{0}_{(1)} + \epsilon^{0}_{(2)}\right) + \dots
\label{318}
\eeq
Inserting this transformation into the (expanded) homogeneity condition,
\beq
A(t, {x})=A^{(0)}(t)+A^{(1)}(t,{x}
)+A^{(2)}(t,{x})=A^{(0)}(\tb),
\label{319}
\eeq
we are then able to fix the corresponding transformation parameters as follows:
\beq
\ep_{(1)}^0=\frac{A^{(1)}}{\dot{A}^{(0)}}, 
~~~~~~
\ep_{(2)}^0=2 \frac{A^{(2)}}{\dot{A}^{(0)}}- \frac{A^{(1)}
\dot{A^{(1)}}}{\left(\dot{A}^{(0)}\right)^2},
~~~~~~
{\ep}_{(1)}^i=0, ~~~~~~ {\ep}_{(1)}=0.
\label{323}
\eeq
Using these particular values into the general transformation laws (\ref{315})--(\ref{317}) immediately 
gives us the transformed quantities $\overline S^{(1)}$, $\overline S^{(2)}$ and $(\sqrt{-\overline{g}})^{(1)}$ in terms of $A$ and of the unbarred fields $S$ and $g$. Inserting such results into Eq. (\ref{310}),  we can finally rewrite the generalized average (or expectation value) of $S$ on the spatial hypersurface at constant $A=A_0$, expanded to second order, in explicit form as 
\bea 
\langle S \rangle_{A_0}&=&  S^{(0)}+\langle S^{(2)} \rangle +
\frac{1}{(\sqrt{-g})^{(0)}}\langle S^{(1)}(\sqrt{-g})^{(1)}\rangle
-\frac{\dot{S}^{(0)}}{\dot{A}^{(0)}(\sqrt{-g})^{(0)}}\langle
A^{(1)}(\sqrt{-g})^{(1)}\rangle \nonumber \\
& -&  \frac{1}{\dot{A}^{(0)}}\left(\langle
A^{(1)}\dot{S}^{(1)}\rangle + \langle \dot{A}^{(1)}S^{(1)}\rangle
\right)+\frac{1}{\dot{A}^{(0)}} 
\left[\frac{\ddot{A}^{(0)}}{\dot{A}^{(0)}}-\pa_t\left(\ln{(\sqrt{-g})^{(0)}}
\right) \right]\langle A^{(1)} S^{(1)} \rangle \nonumber
\\ &+ &  \frac{\dot{S}^{(0)}}{\dot{A}^{(0)}}\left(
\frac{2}{\dot{A}^{(0)}} \langle A^{(1)}\dot{A}^{(1)}\rangle
-\frac{3}{2} \frac{\ddot{A}^{(0)}}{(\dot{A}^{(0)})^2}\langle (A^{(1)})^2
\rangle \right) \nonumber \\ & +&
\left[\frac{\ddot{S}^{(0)}}{2}+
\dot{S}^{(0)} \pa_t\left(\ln{(\sqrt{-g})^{(0)}}
\right)\right]\frac{1}{(\dot{A}^{(0)})^2} \langle
(A^{(1)})^2 \rangle
-\frac{\dot{S}^{(0)}}{\dot{A}^{(0)}} \langle A^{(2)} \rangle. 
\label{324} 
\eea

Here all terms are evaluated at $t=t_0$, and the brackets to the right of this equality denote conventional expectation values or classical averages over all three-dimensional space. We clearly see that the above prescription strongly depends on the scalar observable $A$ chosen to specify the hypersurface to which the averaging is physically referred. However, for any given choice of $A$, we stress that the expression (\ref{324}) is fully independent (up to second order) of the gauge, i.e. $\l \ti S \re_{A_0}= \l  S \re_{A_0}$,
as can be directly checked by applying to $S$, $A$ and $\sqrt{-g}$ a general coordinate transformation (\ref{311}) with arbitrary values of the parameters (see  also the explicit examples of Sect. \ref{sec4}). 

This property of gauge invariance allows to evaluate and compare the  average of a scalar $S(x)$ on different hypersurfaces, defined by  different scalars $A(x)$, by solving the dynamics in a single gauge.
In other words, unlike what is done till now, we do not have to solve a new  dynamics  every time we change the hypersurface of integration.  
We should note, instead, that the result of the conventional average procedure, i.e. $\l S\re= \l S^{(0)}+S^{(1)}+S^{(2)}\re= S^{(0)}+ \l S^{(2)}\re$ (corresponding to the first two terms to the right of the equality sign), is not gauge invariant. 

Equation (\ref{324}) can be rewritten in an equivalent -- but somewhat more convenient -- form as:
\bea 
\langle S \rangle_{A_0}&=&  S^{(0)}+\langle \Delta^{(2)} \rangle +
\frac{1}{(\sqrt{-g})^{(0)}}\langle  \Delta^{(1)}(\sqrt{-g})^{(1)}\rangle
 -  \frac{1}{\dot{A}^{(0)}}\left(\langle
A^{(1)} \Lambda^{(1)}\rangle + \langle \dot{A}^{(1)} \Delta^{(1)}\rangle
\right)  \nonumber \\  &+& \frac{1}{\dot{A}^{(0)}} 
\left[\frac{\ddot{A}^{(0)}}{\dot{A}^{(0)}}-\pa_t\left(\ln{(\sqrt{-g})^{(0)}}
\right) \right]\langle A^{(1)}  \Delta^{(1)} \rangle  +\frac12
\frac{ \Lambda^{(0)}}{(\dot{A}^{(0)})^2} \langle
(A^{(1)})^2 \rangle \, ,
\label{500} 
\eea
where:
\bea
&&
 \Delta^{(i) }=  S^{(i)}- \frac{\dot{S}^{(0)}}{\dot{A}^{(0)}}  A^{(i)} , 
 ~~~~ ~~~~~~ i=1,2, 
 \nonumber \\
 &&
  \Lambda^{(0)} = \ddot{S}^{(0)}- \frac{\dot{S}^{(0)}}{\dot{A}^{(0)}} \ddot{A}^{(0)}, ~~~ ~~~~~~
  \Lambda^{(1)} = \dot{S}^{(1)}- \frac{\dot{S}^{(0)}}{\dot{A}^{(0)}} \dot{A}^{(1)}  .
\label{501} 
\eea
The above formula allows for an easy consistency check. Suppose that $S$ and $A$ are related by an arbitrary function $S = S(A)$. In such a case we should find $\langle S \rangle_{A_0} = S(A_0)$.
Indeed, for $S=S(A)$, inserting in Eq. (\ref{500}) the following (easily derivable) relations,
\bea
&&
 \Delta^{(1) }=0, ~~~~~~~~~~~~~~~~~~~~~~~~ \Delta^{(2) }= \frac12 S''^{(0)} (A^{(1)})^2, 
 \nonumber \\ &&
\Lambda^{(0)} = S^{\prime\prime(0)} (\dot{A}^{(0)})^2, ~~~~~~~~~~~   \Lambda^{(1)}  = S''^{(0)} \dot{A}^{(0)} {A}^{(1)}
\label{502}
\eea 
(a prime denotes the derivative of $S$ with respect to its argument),
only the first term on the r.h.s. survives.

Let us conclude this section by noting that the integration measure in Eq. (\ref{37}) contains the determinant of the full metric $\overline g_{\mu\nu}$ of ${\cal M}_4$, and not the determinant $\overline\gamma \equiv \det (\overline\gamma_{i j})$ of the intrinsic metric $\overline\ga_{ij}= \overline{g}_{ij}$ of the hypersurface $\Sg_A$. This represents a further difference between our prescription and other averages prescriptions commonly used in the literature, using the weight $\sqrt{|\overline\ga |}$ for the integration over the hypersurfaces determined by the chosen space-time foliation (see e.g. \cite{3,6}).

It may be noted that a modified averaging procedure with
$|\overline{\ga}|$ replacing $-\overline{g}$ in Eq. (\ref{37}) is still compatible with gauge invariance: indeed, such a modification can be obtained from a covariant four-dimensional integral simply by replacing $\da(A-A_0)$, in the window (\ref{32}), with the following -- more complicated but still covariant -- window function:
\beq
\da(A(x)-A_0) \sqrt{\left|g^{\mu\nu} \pa_\mu A\pa_\nu A\right|}.
\label{325}
\eeq
In that case, by repeating the same procedure as before, and using the relation $\overline g^{00}\, \overline g= \overline\ga$, we end up with a result which can be written exactly as in Eq. (\ref{37}), but with  $-\overline{g}$ replaced by  $|\overline{\ga}|$. 

It should be stressed that even in that case a consistent gauge-invariant prescription implicitly requires a coordinate transformation (to the barred frame) of the variables to be averaged. In that case, after  expressing  the barred variables in terms of the unbarred ones, the average prescription can be written in explicit second-order form as 
\bea 
\langle S \rangle_{A_0}&=&  S^{(0)}+\langle \Delta^{(2)} \rangle +
\frac{1}{\left(\sqrt{|\gamma|}\right)^{(0)}}\langle \Delta^{(1)}\left(\sqrt{|\gamma|}\right)^{(1)} \rangle
 -  \frac{1}{\dot{A}^{(0)}} \langle
A^{(1)}\Lambda^{(1)}\rangle  \nonumber \\  &-& \frac{1}{\dot{A}^{(0)}} 
\pa_t\left(\ln \left(\sqrt{|\gamma|}\right)^{(0)}\right) \langle A^{(1)} \Delta^{(1)} \rangle  +
\frac12 \frac{ \Lambda^{(0)}}{(\dot{A}^{(0)})^2} \langle
(A^{(1)})^2 \rangle 
\label{505} 
\eea
(where, to this order, we can use $\ga= \det (g_{ij})$), 
replacing Eq. (\ref{500}) for the case of the window function (\ref{325}).
In this convenient form the same consistency checks discussed for (\ref{500}) can be easily repeated. 

The two averaging prescriptions (using, respectively, $\overline g$ or $\overline \ga$ in Eq. (\ref{37})) are clearly inequivalent in all cases where $\sqrt{-\overline g}/ \sqrt{|\overline\ga|}$ is non-homogeneous.  In the rest of this paper we will concentrate our discussion on the -- apparently more natural, in a covariant four-dimensional context -- prescription defined in terms of the full metric $g_{\mu\nu}$, and corresponding to Eqs. (\ref{324}), (\ref{500}).

\section {Examples of expectation values for the quantum \\ backreaction}
\label{sec4}
\setcounter{equation}{0}

It is well known that the computation of backreaction effects, due to the presence of small inhomogeneities perturbing the large-scale cosmological evolution, is plagued by substantial ambiguities arising from the gauge dependence of the perturbative approach and of the adopted averaging procedure. For the backreaction of the quantum vacuum fluctuations in inflationary backgrounds, in particular, there is a controversial literature (see e.g. \cite{10}--\cite{13}), where different results are obtained on the grounds of computations performed with different methods in different gauges. 

The aim of this section is present a few simple explicit examples showing that, differently from the conventional procedures used in the literature, our generalized average prescription (\ref{324}) always gives gauge-invariant results (up to second perturbative order), thus avoiding interpretation problems and preventing ambiguous conclusions. 

We shall work in the context of a spatially flat, FRW background geometry, sourced by a single scalar field $\phi$ according to the Einstein equations, and we expand our background fields $\{\phi$, $g_{\mu\nu}\}$ up to second order in the non-homogeneous perturbations, without fixing any gauge, as follows:
\bea
&&
\phi(t,\vec{x})=\phi^{(0)}(t)+\phi^{(1)}(t,\vec{x})+\phi^{(2)}(t,\vec{x}),
\label{41} \\ 
& & g_{00}= -1-2 \a^{(1)}-2 \a^{(2)}, ~~~~~~~\,\,\,\,\,\,\,\,\,\,
g_{i0}=-{a\over2}\left(\beta^{(1)}_{,i}+B^{(1)}_i\right) 
-{a\over2}\left(\beta^{(2)}_{,i}+B^{(2)}_i\right)\,,
\nonumber \\
& & 
g_{ij} = a^2 \left[ \delta_{ij} \left(1-2 \psi^{(1)}-2 \psi^{(2)}\right)
+D_{ij} (E^{(1)}+E^{(2)}) +
{1\over 2} \left(\chi^{(1)}_{i,j}+\chi^{(1)}_{j,i}+h^{(1)}_{ij}\right) \right.
\nonumber \\
& & \left.
\,\,\,\,\,\,\,\,\,\,\,\,\,\,+
{1\over 2} \left(\chi^{(2)}_{i,j}+\chi^{(2)}_{j,i}+h^{(2)}_{ij}\right)\right],
\label{42}
\end{eqnarray}
where $D_{ij}=\partial_i \partial_j- \delta_{ij} (\nabla^2/3)$, and $a=a(t)$ is the scale factor of the homogeneous FRW  metric. Here 
 $\a^{(1)}$, $\b^{(1)}$, $\psi^{(1)}$, $E^{(1)}$ are pure scalar first-order perturbations, 
$B^{(1)}_i$ and $\chi^{(1)}_i$ are transverse vectors ($\pa^i B^{(1)}_i=0$ and 
$\pa^i \chi^{(1)}_i=0$), $h^{(1)}_{ij}$ is a traceless and transverse tensor 
($ \pa^i h^{(1)}_{ij}=0=h^{(1)\,i}_i$), and the same notation applies to  the case of the second-order perturbations. 

Let us start with the (almost trivial, but instructive) example concerning the average of the scalar field $\phi$, using as our family of reference  hypersurfaces the ones where $\phi$ itself takes constant values. Let us then apply our average prescription (\ref{324}) with $S(x)= \phi(x)$ and $A(x)= \phi(x)$, on the hypersurface where $A_0= \phi^{(0)}(t_0)$. 
This is a particular example of the general case $S=S(A)$ discussed in  Eq. (\ref{502}) and it is clear, already from an intuitive point of view, that the average should yield the result:
\beq
\langle \phi(t,\vec{x}) \rangle_{\phi_0}=\phi^{(0)}(t_0). 
\label{43}
\eeq

This is indeed the case, as can be quickly checked by computing the average (\ref{324}) in the so-called Uniform Field Gauge (UFG) fixed by the condition $\phi^{(1)}=0=\phi^{(2)}$, i.e. the gauge where the scalar field is homogeneous. In that case, in fact, all terms on the r.h.s. of Eq. (\ref{324}) are identically vanishing except the first one, and the result (\ref{43}) is trivially obtained. But the same result is obtained in any other gauge, even if $\phi$ is not homogeneus, and 
$\phi^{(1)}$, $\phi^{(2)}$ are both nonvanishing: putting $S=\phi=A$, in fact, we find that there is an exact cancellation among all terms on the r.h.s. of Eq. (\ref{324}) except the first one, and we are led again to the result (\ref{43}). 

The situation is different if we consider instead the conventional average prescription, corresponding to the first two terms on the r.h.s. of Eq. (\ref{324}), namely to 
\beq
\langle \phi(t,\vec{x}) \rangle=\phi^{(0)}(t)
+\langle \phi^{(2)}(t,\vec{x}) \rangle .
\label{44}
\eeq
In such a case we can obtain different results in different gauges. 
 In the following we shall reserve the notation $ \langle \dots \rangle_{A_0}$ for our gauge invariant prescription and  $\langle \dots \rangle_{g_i}$ for the conventional quantum average in  the gauge $g_i$.

In the particular case of the UFG, for instance, we have $\phi^{(2)}=0$ and we exactly recover the previous result (\ref{43}) for any given model of background evolution. But in other gauges the correction to the background value $\phi^{(0)}$ can be nonzero. We may consider, for instance, the so-called Uniform Curvature Gauge (UCG) (also called off-diagonal gauge \cite{14}), where we set to zero the variables $\psi$ and $E$ of Eq. (\ref{42}) (see also \cite{15} for a general discussion of various possible gauges). In that gauge $\phi^{(2)}\ne0$, in general, as also confirmed by previous studies of the backreaction in models of chaotic inflation \cite{12} with scalar-field potential $V= m^2 \phi^2/2$. For such models, the computation of the vacuum expectation value (v.e.v.) of the quantum fluctuations of the scalar inflaton field in the UCG (neglecting vector and tensor perturbations to first order), expanded in the long-wavelength limit, and to leading order in the slow-roll parameter $\ep$, gives, according to Eq. (\ref{44}), 
\beq
\langle \phi(t,\vec{x}) \rangle_{UCG} = \phi^{(0)}(t) \left[1-
\frac{\epsilon}{4 M_{\rm P}^2} \langle \phi^{(1)\,2}  \rangle\right] 
\label{45}
\eeq
($M_{\rm P}$ is the Planck mass, and $\ep= - \dot H/H$, where $H = \dot a /a$ ). 
This is clearly different from the UFG result $\langle \phi(t,\vec{x}) \rangle_{UFG} = \phi^{(0)}(t)$, valid for the same model of chaotic inflation to all orders in $\ep$. There is nothing wrong, of course, with such a discrepancy, since the physical meaning of $\langle \phi(t,\vec{x}) \rangle$ is different in different gauges  ($\phi$ is not a gauge-invariant perturbation even at first order).

As a second example we will consider  the average of a scalar
quantity playing a central role in the computation of the inflationary backreaction \cite{11}-\cite{13}: the so-called volume expansion $\Theta$, determined by the covariant divergence of a bundle of comoving curves. For a scalar-field-dominated geometry, in particular, we have 
\beq
\Theta=\nabla_\mu u^\mu, ~~~~~ ~~~~~~~
u_{\mu}=\frac{\partial_\mu \phi}{\left(-g^{\alpha \beta}\partial_\alpha \phi 
\partial_\beta \phi\right)^{1/2}} \,, 
\label{46}
\eeq
so that $\Theta= 3H$ for the unperturbed homogeneous geometry. 

If we include perturbations, up to second order as in Eq. (\ref{41}), (\ref{42}), and we compute the  average of $\Theta$ according to the standard prescription, the result is notoriously gauge dependent. Consider, for instance, the previous model of chaotic inflation: including the contribution of non-homogeneous quantum fluctuations (to second order), and working in the UFG, one finds for the v.e.v. of $\Theta$ no correction to the background value to first order in the slow-roll parameter $\ep$, namely:
\beq
\langle \Theta \rangle_{UFG}=3 H \left[1+ {\cal O}(\epsilon^2)\right] 
\label{47}
\eeq
(see also  \cite{10,11} for similar UFG results). 
Repeating the computation in the UCG one finds \cite{12}, instead, a nonvanishing backreaction even to first order in $\ep$:
\beq
\langle \Theta \rangle_{UCG}=3 H \left[1-{\epsilon \over 4}
\frac{\langle \phi^{(1)\,2} \rangle}{M_{\rm P}^2}+ {\cal O}(\epsilon^2)\right]. 
\label{48}
\eeq
Note that, in this gauge, $ \phi^{(1)}$ also corresponds to the canonical, gauge-invariant Mukhanov variable for scalar perturbations \cite{17}.

This difficulty (and possible source of misleading conclusions about the backreaction of quantum fluctuations in chaotic models) is absent if we apply our prescription (\ref{324}), which gives the same result in any gauge. If we refer, in particular, such a covariant average to the convenient hypersurface at constant inflaton (i.e. putting $A=\phi$ as before), we find that there is no backreaction to first order in $\ep$, 
\beq
\langle \Theta \rangle_{\phi_0}=3 H \left[1+ {\cal O}(\epsilon^2)\right]. 
\label{49}
\eeq
(in agreement with the UFG result of the conventional v.e.v. prescription). 
We stress that the above result is insensitive to the choice of the gauge but not insensitive -- in principle -- to the choice of the reference hypersurface. 

The two examples presented up to now might suggest that the result of a conventional average, when performed in the UFG, should coincide with that of the gauge invariant procedure  performed with $A=\phi$. More generally, one might even think that the result obtained in a gauge where a given scalar variable is homogeneous should coincide with the result of the gauge-invariant average performed on the hypersurface where the same variable is constant. 
This, however, is not in general true. Considering the choice $A=\phi$, for instance, we find that the covariant average $\l S \re_{\phi_{o}}$ reduces in the UFG to the first three terms on the r.h.s. of Eq. (\ref{324}): hence, it differs from the conventional average $\l S\re$ -- corresponding to the first two terms on the r.h.s. -- by the contribution of  $\langle S^{(1)} (\sqrt{-g})^{(1)}\rangle$. Only when this term is zero the two results coincide. 

For a further illustration of this point we may recall that, in the context of the UFG choice, we are free to fix the gauge by imposing another condition on the scalar components of the metric perturbations: in particular, we can set to zero one scalar variable among $\b$, $\psi$ and $E$ of Eq. (\ref{42}) (but not $\a$, or the background would become trivially homogeneous up to first order, see \cite{16}). These three different possibilities identify three possible UFGs, connected among them by a coordinate transformation of the type  (\ref{311}) with $\epsilon_{(1)}^0=\epsilon_{(2)}^0=0$ (namely, by a  transformation leaving the time parameter unchanged). Under such a transformation $S^{(1)}$ is unchanged, but not $S^{(2)}$ and $ (\sqrt{-g})^{(1)}$ (see Eqs. (\ref{315})--(\ref{317})). Hence, going from one UFG choice to another, the conventional average $\l S \re$ (only depending on $\l S^{(2)} \re$) may change, with a corresponding change of the term $\langle S^{(1)} (\sqrt{-g})^{(1)}\rangle$, controlling the difference with the gauge invariant average (obviously, the sum of the two terms, determining  the gauge-invariant combination $\l S \re_{\phi_{o}}$, is left unchanged). 

It seems appropriate to include here an explicit example illustrating the differences among the various UFG choices. We shall report  the results of a computation  \cite{16} of $\l \Theta \re$ for a cosmological background perturbed around a power-law geometry with scale factor $a(t) \sim |t|^{1/3}$ (solution of the Einstein equations with vanishing scalar potential).  Working in the UFG fixed by $\b^{(1)}=0= \b^{(2)}$ one obtains: 
\beq
\langle \Theta \rangle_{UFG \b}=
3 H \left[ 1 +\frac{45}{8} {\l  Q^{(1)\,2}\re_{\rm REN}\over M_{\rm P}^2} \right],
\label{410}
\eeq
while in the UFG with $E^{(1)}=0= E^{(2)}$ one obtains:
\beq
\langle \Theta \rangle_{UFG E}=
3 H \left[ 1 -\frac{3}{4} {\l  Q^{(1)\,2}\re_{\rm REN}\over M_{\rm P}^2} \right]. 
\label{411}
\eeq
Here $Q$ denotes the gauge-invariant Mukhanov variable, and the suffix REN denotes that the v.e.v. has been regularized through a suitable adiabatic subtraction. These two results not only differ between each other, but also differ from the one obtained with the gauge invariant prescription (\ref{324}) with $A=\phi$, which gives
\beq
\langle \Theta \rangle_{\phi_0}=
3 H \left[ 1 -{15\over 8} {\l Q^{(1)\,2} \re_{\rm REN}\over M_{\rm P}^2} \right]
\label{412}
\eeq
in any gauge.

It follows, in particular, that we cannot 
try to solve the problem of the gauge dependence of the backreaction 
considering the UFG as a privileged gauge, as often suggested in the literature. The need for a truly gauge  invariant approach is evident from the above example but, for a better understanding of its importance, let us conclude the section with a few comments stressing the different effects of gauge transformations on different types of  average prescriptions. We shall compare, in particular, our approach with the one more frequently used in the literature.  

According to the approach proposed in this paper the averaging procedure is
covariantly defined, on the hypersurface where a suitably given scalar field
$A(x)$ is constant, in terms of fields ($S$, $g_{\mu\nu}$, \dots) evaluated in
the coordinate frame (or gauge) where $A$ is homogeneous (the barred frame,
following the notations of Sect. \ref{sec3}). Only in this frame -- i.e. only
in terms of the barred fields -- our average can be written in the simple
classical or quantum form given, respectively, by Eq. (\ref{37}) or
(\ref{38}). If we change gauge, namely if we want to rewrite the average in
terms of our fields evaluated in a different coordinate frame where $A$ is no
longer homogeneous, then our prescription assumes a new, more complicated form
if expressed through the conventional definition of volume averages (or
expectation values), see Eq.(\ref{324}). However, such a new form is always covariantly related (by definition) to the original prescription given in Eq. (\ref{37}) (or (\ref{38})), and this definitively explains why, at the end of our computations, the same result is obtained in any gauge. 

According to the more conventional approaches appearing in the  literature,
instead, when one changes the gauge -- i.e. when one moves from the  frame
where the scalar field $A(x)$ is homogeneous to the frame where, for example,
another variable (or a set of not necessarily scalar variables)
is homogeneous -- one still maintains the same definition of  spatial average
(or expectation value) used in the old gauge, and simply expresses  it in terms
of the new fields, i.e. of the fields trasformed in the new frame.  If the
original prescription is covariantly defined, and if one of the  variables
imposed to be homogeneous in the new gauge, say $Z(x)$,  is a scalar, then the
new average prescription can be interpreted  as the same as the previous one,
performed however on a different  hypersurface: the one where $Z$ is constant
(for example, we can interpret in this way the results in \cite{3}).
But if there is no homogeneous scalar variable available in the new  gauge,
and/or the average is not covariantly defined, then changing  the gauge
implies a true modification of the averaging prescription  (not equivalent to
a change of the reference hypersurface). In both  cases it is clear that,
within this conventional approach,  different gauges lead to different results
for the averaging  procedure.

\section{Summary}
\label{sec5}
\setcounter{equation}{0}

Let us briefly summarize our main results and conclusions.
It is generally acknowledged that, in order to extract physically meaningful
results from cosmological perturbations, it is necessary to eliminate all
possible  gauge artifacts. At linear order in the perturbations, the way to do
this is well known:  following Bardeen's pioneering work \cite{Bardeen}, it
consists of defining gauge invariant combinations of the perturbations
themselves. Such gauge invariant combinations  have a clear physical meaning,
typically representing the perturbation of various quantities on hypersurfaces
along which some given variable is constant: they are like  ``relative''
fluctuations of different variables.

A more complicated problem arises when perturbations are
considered to second order. This can be necessary for a variety of reasons: i)
when the perturbations are too large to justify the leading order
approximation; ii) when the  leading-order result vanishes (e.g. for
non-gaussianity in simple inflationary model); or, finally, iii) when one is
interested in evaluating backreaction effects either at the classical or at
the quantum level.
 
The procedure followed at first order  can be generalized to define gauge 
invariant combinations of the second order  perturbations  (see, for
example, \cite{NGexample}). 
However,  whenever one has to deal with averages, like in the issue  of   
backreaction effects discussed in this paper, such second-order
gauge  invariant variables are not of much use.

For the cosmological backreaction, in particular, a simplification 
occurs due to the fact that we are interested in spatial (or quantum) averages 
of second-order perturbations at very large scales.
By carefully studying the gauge-transformation properties of four-dimensional integrals (while stressing the distinction and relation between gauge and general coordinate transformations) and by then going to the limit of an infinitesimal time-like shell, we have been able to propose a general formula for the
classical or quantum average of any scalar quantity, on hypersurfaces on which another given scalar quantity is constant. Our non-trivial proposal, Eq.(\ref{324}), can be shown to be gauge-invariant in the quantum case and in the classical one for very large spatial-averaging volumes.

Our proposal is at variance with the recipes that can be found in the  literature (often with contrasting results and conclusions) since, in {\em any}  gauge, it contains important extra terms. Neglecting those terms  induces, in general, gauge-invariance violations  that  are  of the  same order as the effect one wishes to estimate. This may well  explain some of the disagreement that can be found in the current  literature.

Unfortunately, the requirement of gauge invariance alone does not fix uniquely our    prescription. Our simplest gauge-invariant recipe  involves the determinant of the four-dimensional metric and comes 
naturally from taking the thin-time-slice limit of a gauge-invariant 
four-dimensional integral. It is also possible to define another
gauge-invariant average (using Eq. (\ref{325})), which, in adapted coordinates, reduces to a more conventional integration  measure
involving the induced metric on the hypersurface on which the average is computed. While the choice between one or the other definition should be dictated by  the physics of the problem, in both cases the proposed gauge-invariant averages differ in a non-trivial way from the ones that are  currently used.

In this paper we have only given the general ideas behind our proposal, have checked its consistency in a number of cases, and discussed some simple applications. A more detailed discussion of its implications in specific issues is deferred to further work.

\section*{Acknowledgements}

It is a pleasure to thank for their warm hospitality and their support the following scientific institutes, where part of this work was performed: 
the Galileo Galilei Institute for Theoretical Physics (Florence), the Theory Unit of the Physics Department of CERN (Geneva), and the Department of Physics of the University of Bologna. One of us (GV) is also grateful to G. Vilkovisky for useful correspondence.

\end{document}